# Nonlinear Characterization of Thin-Film $LiNbO_3$ Acoustic Filters

Omar Barrera*, Bryan T. Bosworth*, Taran Anusorn*, Kenny Huynh, Ian Anderson, Nicholas R. Jungwirth, Michael Liao, Sinwoo Cho, Jack Kramer, Lezli Matto, Mark S. Goorsky, Nathan D. Orloff, and Ruochen Lu

***Abstract*—Compact, high-performance components in millimeter-wave (mmWave) communication systems demand new acoustic filter technology at increasingly higher frequencies. Among various promising mmWave platforms, first-order antisymmetric (A1) mode laterally excited bulk acoustic resonators (XBARs) in thin-film lithium niobate ($LiNbO_3$) have perhaps the most impressive linear performance. Despite these advances, there are few reports of nonlinear characterization of $LiNbO_3$ filters at mmWaves. Here, we address this gap by developing a new nonlinear methodology for high-frequency filters. The result is a methodology for performing power-dependent S-parameters and third-order intermodulation (IMD3) measurements. To test our methodology, we fabricated filters on transferred single-crystal $LiNbO_3$ films on sapphire ($Al_2O_3$) and silicon (Si) substrates with amorphous silicon (aSi) sacrificial layer. At 21.8 GHz, the filters on $Al_2O_3$ demonstrated an insertion loss of 1.48 dB, a 3 dB fractional bandwidth (FBW) of 17.7%, and in-band third-order input intercept points (IIP3) of 50.8 dBm. At 21.6 GHz, the filters on silicon demonstrated an insertion loss of 2.47 dB, a 3 dB FBW of 18.6%, and in-band IIP3 of 46.5 dBm. The nonlinear results conclusively show that thermal stability and passband distortion improved on the $Al_2O_3$ substrate, confirming that substrate selection plays a pivotal role in mitigating nonlinearity in acoustic front-end modules.**

## I. INTRODUCTION

COMMERCIAL wireless systems rely on acoustic filters as the preferred solution for compact and efficient RF front-end (RFFE) filtering in sub-6 GHz frequency bands [1], [2], [3]. Surface acoustic wave (SAW) and bulk acoustic wave (BAW) technologies dominate the market because they offer high integration density and a small footprint, owing to the short wavelengths of acoustic waves relative to their electromagnetic (EM) counterparts [4]. As wireless standards evolve and demand ever-increasing data rates and spectral efficiency, many researchers expect acoustic filters to remain the backbone of current 5G deployments, future 6G systems, and beyond [5], [6]. With this progression to the millimeter-wave (mmWave) spectrum, there is a growing need for high-performance acoustic filter platforms that can operate efficiently at much higher frequencies [7].

This paper is an expanded version of the IEEE International Conference on Micro Electro Mechanical Systems (MEMS) 2024.

Manuscript received XX; revised XX. This work was supported by the Defense Advanced Research Projects Agency (DARPA) Compact Front-End Filters at the Element-Level (COFFEE) program and NSF CAREER 2339731.

O. Barrera, T. Anusorn, I. Anderson, S. Cho, J. Kramer, and R. Lu are with Chandra Family Department Electrical and Computer Engineering, The University of Texas at Austin, Austin, TX, 78712, USA (email: taran.anusorn@utexas.edu).

B. T. Bosworth, N. R. Jungwirth, and N. D. Orloff are with the National Institute of Standards and Technology (NIST), Boulder, CO, 80305, USA.

K. Huynh, M. Liao, L. Matto, and M. S. Goorsky are with the University of California, Los Angeles, CA, 90095, USA.

O. Barrera, B. T. Bosworth, and T. Anusorn contributed equally.

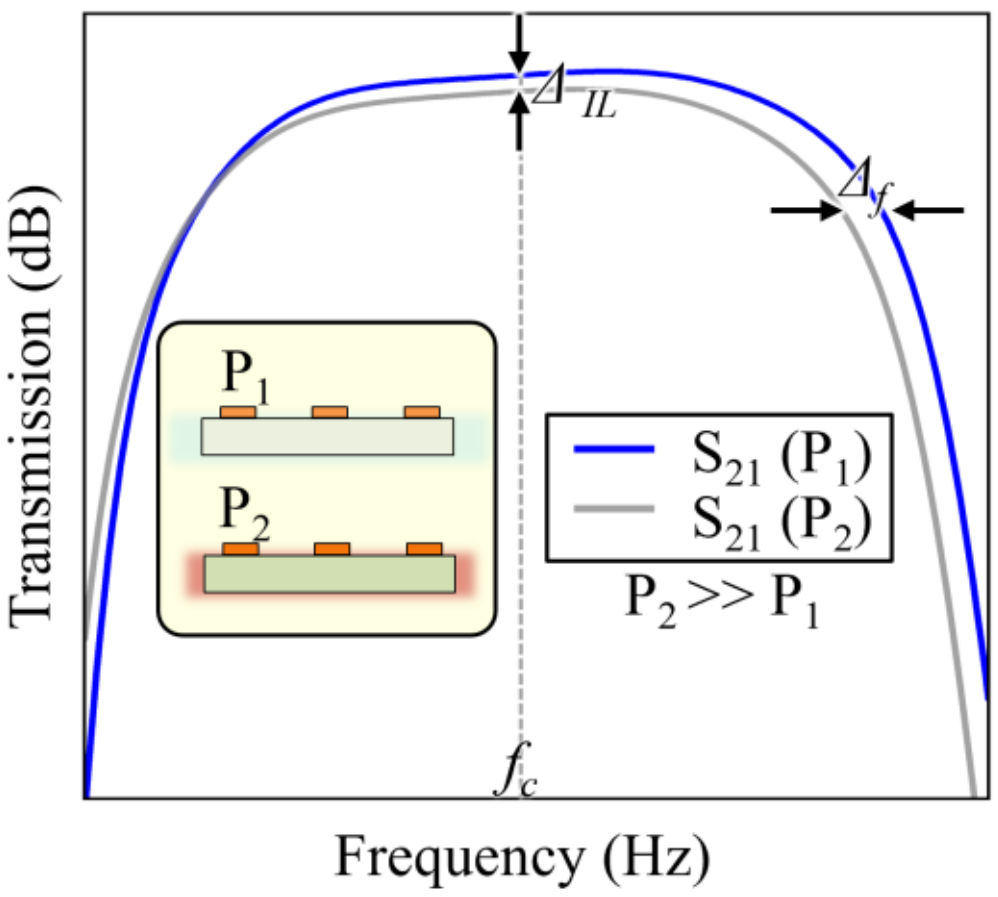

**Fig. 1.** Depiction of the filter passband shift under different power (P) conditions, i.e., $P_2 >> P_1$. For a suspended laterally excited bulk acoustic resonator (XBAR), thermally induced nonlinearity causes an increase in insertion loss (IL) and a shift in the passband frequency. An inset illustrates thermal effect in XBAR devices under different power levels. $f_c$ represents the center frequency of the filters, while $\Delta f$.

Scaling the operating frequency of acoustic resonators is conceptually straightforward. SAW devices achieve this by reducing the size of the interdigitated electrode, while BAW devices rely on thinner piezoelectric films [8], [9]. However, both approaches face significant practical challenges. SAW devices encounter lithographic limitations and increased resistive losses as electrode width decreases. Meanwhile, BAW resonators suffer from increased acoustic losses due to material degradation in the ultra-thin-film regime [10] and parasitic capacitance that can overwhelm the device performance. Despite these issues, recent academic efforts have demonstrated acoustic resonators and filters operating well above 10 GHz, extending into the mmWave regime. These efforts employed advanced piezoelectric materials, including aluminum nitride (AlN) [11], scandium-doped AlN (ScAlN) [12]–[22], and lithium niobate ($LiNbO_3$) [23], [24].

As for $LiNbO_3$, the platform is a strong contender for mmWave acoustic filter technology [25]. $LiNbO_3$ has a high $e_{15}$ piezoelectric coefficient that is compatible with the lateral excitation of acoustic modes [26]–[28]. Such properties led to the development of first-order antisymmetric (A1) mode laterally excited bulk acoustic resonators (XBARs). By transferring a single crystal 100-nm-thick $LiNbO_3$ onto low-loss substrates such as sapphire ($Al_2O_3$) or silicon (Si) with an amorphous silicon (aSi) bonding and sacrificial layer, researchers demonstrated resonators with high quality factor ($Q$) and large electromechanical coupling ($k^2$) [29], [30], two critical metrics for acoustic resonators towards high-performing filters. These advances mark a significant step towards commercial compact, high-performance acoustic filters [31]–[34] that operate well in the milliwatt range [35]. Despite this progress, most researchers focus on linear characterization of filter behavior, leaving their nonlinear performance, which is crucial for high-power and high-dynamic-range applications, unexplored.

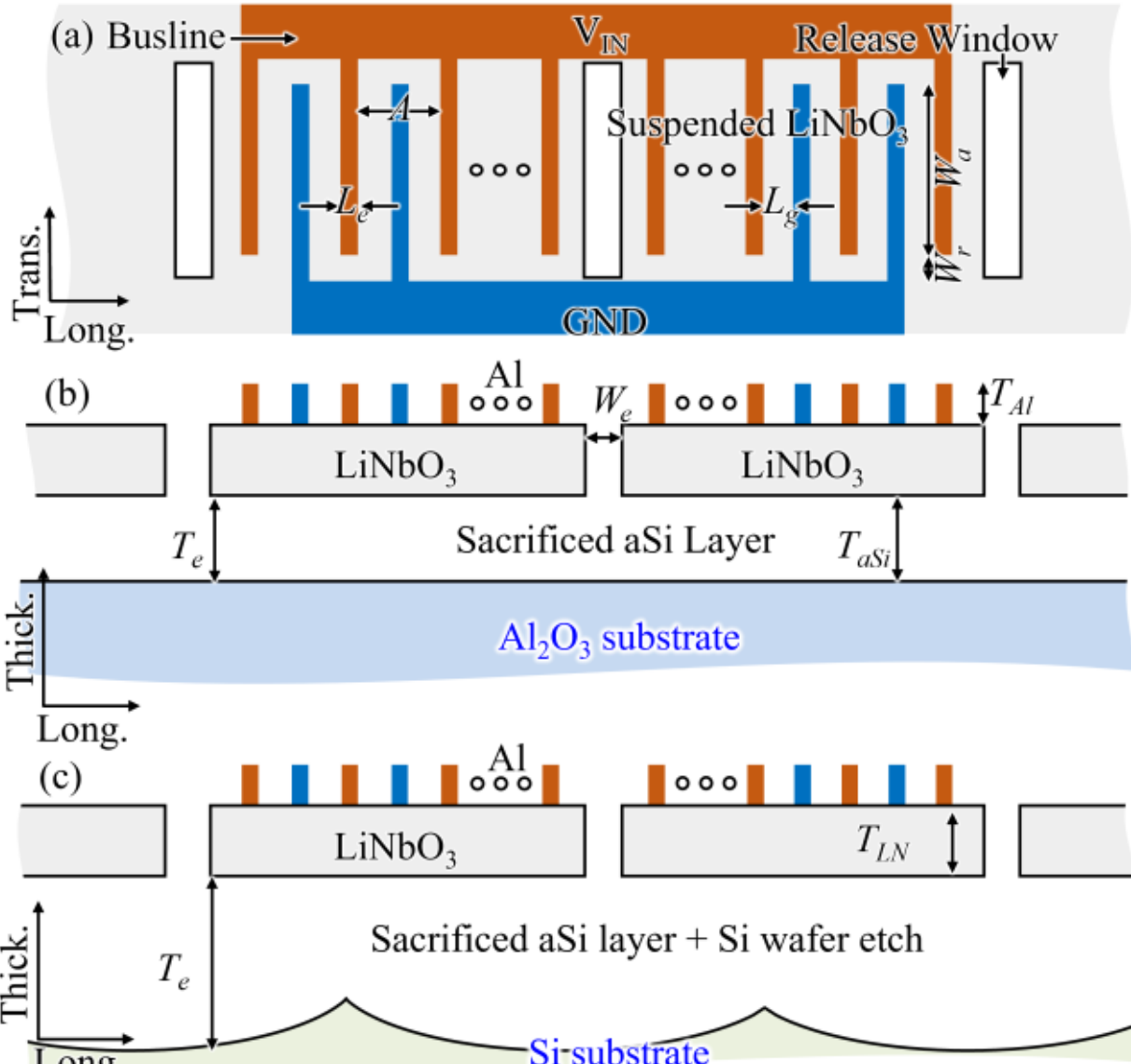

**Fig. 2.** (a) XBAR resonator top schematic. (b) Cross-sectional view of released XBAR resonator in $LiNbO_3$-aSi-Al2$O_3$, where aSi bonding layer is sacrificed and an air cavity with defined thickness is formed. (c) Cross-sectional view of released XBAR in $LiNbO_3$-aSi-Si, where aSi layer and Si carrier are isotropically etched, creating a larger air cavity.

Nonlinear filter metrology is important. At high power levels, acoustic resonators exhibit nonlinear behavior that can significantly degrade system performance. A primary contributor is self-heating, in which elevated input power raises the resonator's temperature, thereby shifting its frequency response due to temperature-dependent material properties [36]. $LiNbO_3$ is known for its low thermal conductivity of 5.8 W/(m·K) and high temperature coefficient of frequency (TCF) around −80 ppm/K [37] for the A1 mode [38], [39]. The low temperature stability, as well as the thermal isolation from the suspended thin-film structure of mmWave XBARs, makes it especially susceptible to such frequency drift under high-power operation (see Fig. 1). In addition to distorted filter performance, nonlinear mixing effects become more pronounced at elevated powers, generating harmonics and intermodulation products [40], [41], especially the third-order intermodulation (IMD3) [42]. These spurious signals can desensitize RFFEs, compromising signal integrity and limiting the linearity of communication systems. Similarly, the IMD3 has not been well studied in mmWave XBARs.

In the following, we investigate two closely related stacks of $LiNbO_3$-based materials on different carrier substrates: Si and $Al_2O_3$. We present a comprehensive characterization of their material properties, linear performance, and, notably, nonlinear behavior. Nonlinear measurements, including high-power S-parameter testing and in-band third-order input intercept points (IIP3), assess linearity across both platforms. These results are particularly important for evaluating platforms for applications that require high linearity, such as high-power or densely packed RFFEs.

This work is the journal extension of a conference proceeding [1], incorporating new device data, in-depth analysis, and, most importantly, an emphasis on the nonlinear characterization of mmWave XBAR filters. The organization of the paper is as follows. Section II describes the XBAR filter testbed design and simulated linear performance. Section III presents thermal simulations comparing nonlinear heating in $LiNbO_3$-aSi-$Al_2O_3$ and $LiNbO_3$-aSi-Si platforms. Section IV provides material analysis using high-resolution X-ray diffraction (HRXRD) and transmission electron microscopy (TEM). Section V details device fabrication and linear measurement results. Section VI introduces the nonlinear measurement system. Section VII discusses power-dependent S-parameter measurements. Section VIII reports IMD3 measurements. Section IX compares the results to the state of the art (SoA). Section X concludes the paper.

TABLE I
DESIGN PARAMETERS OF SERIES AND SHUNT RESONATORS IN THE XBAR FILTER TESTBED. PARAMETER SYMBOLS ARE DEFINED IN FIG. 2

| Sym. | Parameter | Value | |
|---|---|---|---|
| | | Shunt | Series |
| $A$ | Cell length (μm) | 6 | |
| $L_e$ | Electrode length (μm) | 0.8 | |
| $L_g$ | Gap length (μm) | 2.2 | |
| $W_a$ | Aperture width (μm) | 71 | |
| $W_r$ | Recessed width (μm) | 6 | |
| $W_e$ | Etch window width (μm) | 5 | |
| $T_{LN}$ | $LiNbO_3$ thickness (nm) | 110 | |
| $T_{aSi}$ | Sacrificial aSi thickness (nm) | 1000 | |
| $T_{Al}$ | Aluminum thickness (nm) | 350 | |
| $T_e$ | Etch release extension (μm) | 1 ($LiNbO_3$-aSi-$Al_2O_3$)/ 35 ($LiNbO_3$-aSi-Si) | |
| $N$ | No. of finger pairs | 24 | 11 |

## II. XBAR FILTER DESIGN AND SIMULATION

Our experiment required two different but very similar instantiations of what was nominally the same resonator. The design approach was identical for both Si and $Al_2O_3$ carrier substrates. From the top view [Fig. 2(a)], the active region of the resonator consists of interdigitated transducers (IDTs) on a thin-film $LiNbO_3$ piezoelectric layer [Fig. 2]. From the side view [Fig. 2(b) and (c)], the thickness-shear A1 mode by IDTs via the $e_{15}$ piezoelectric coefficient is illustrated [25], [34]. In both cases, the electrodes are aluminum with a thickness of 350 nm. The spacing between electrodes is 3 μm, in accordance with the guidelines reported in [30].

After release [Figs. 2(b) and 2(c)], the anchors connected the resonators to the substrate. Both stacks contained a 1 μm aSi layer, serving as both the bonding and sacrificial layer. The difference between the two stacks lay in the architecture of the release cavities and the resulting thermal behavior. On the $LiNbO_3$-aSi-$Al_2O_3$ platform, xenon difluoride ($XeF_2$) etching step completely removed the sacrificial aSi layer, forming a well-defined vertical air cavity. On the $LiNbO_3$-aSi-Si platform, the $XeF_2$ etching step resulted in a much deeper air-gap $T_e$ because both the aSi layer and the underlying Si carrier were isotropically etched. Notably, uneven vertical cavity profile was formed in the latter case following the locations of its release windows. For this reason, a nominal depth $T_e$ = 35 μm was assumed in this case for simplicity. Table I shows a summary of the resonator dimensions, along with the thermal properties of each substrate. The difference in vertical cavity sizes, and thus effective cavity volumes, directly affected thermal isolation and heat dissipation. As a result, the two resonators were expected to exhibit significantly different nonlinearities under high-power operation.

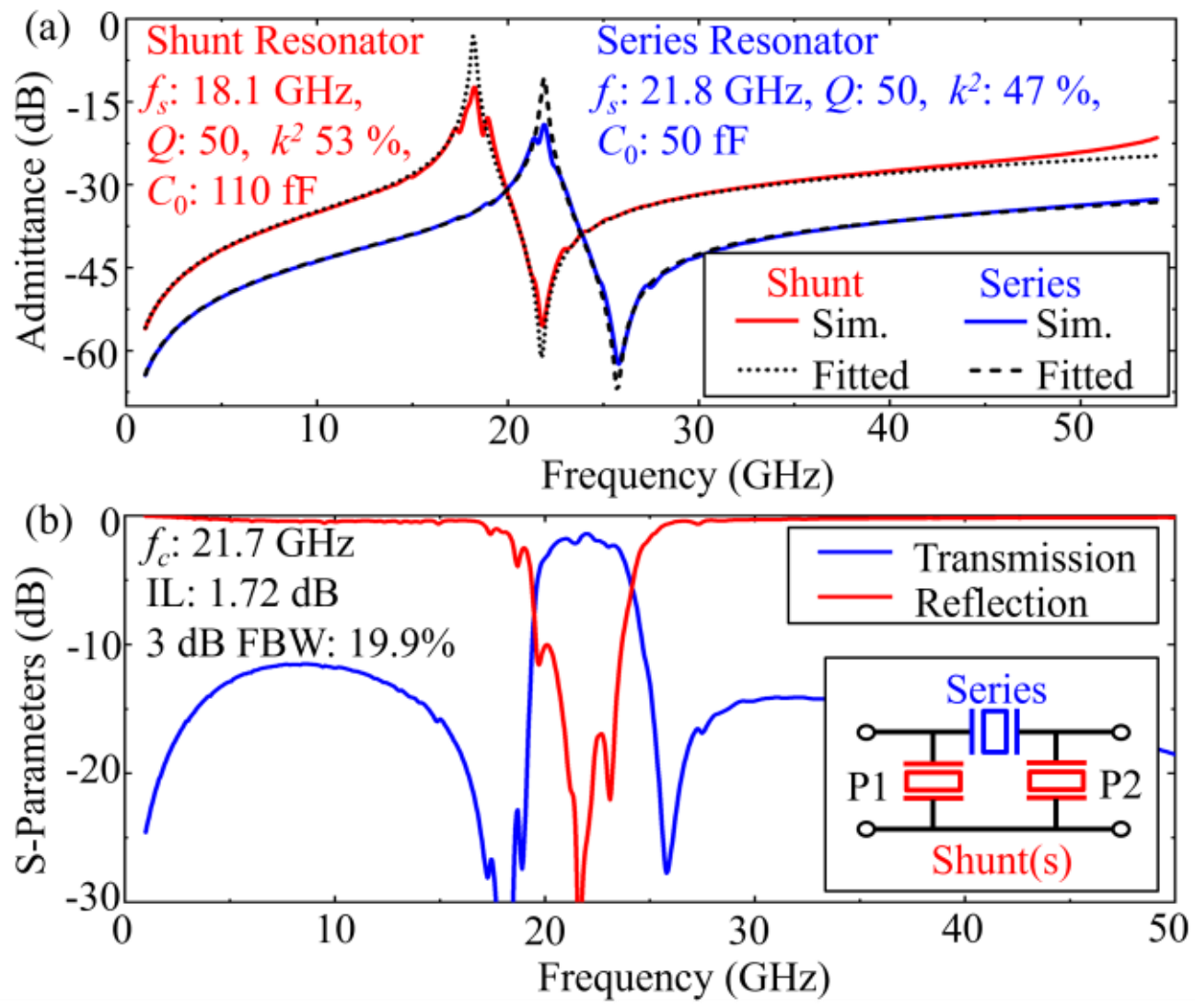

**Fig. 3.** (a) FEA simulated XBAR admittance with extracted key modified Butterworth-Van Dyke (mBVD) fitted performance specifications. (b) Simulated third-order ladder filter response. Ports 1 and 2 of the filters are indicated as P1 and P2.

We used COMSOL[1] Multiphysics finite element analysis (FEA) to simulate the series and shunt resonators [Fig. 3(a)]. The detailed simulation procedure, acoustic mode shape, and key parameter extraction have been well reported in [1]. It should be noted that the spurious responses observed in the simulated results are likely attributed to the interdigital transducer (IDT) metallization and nonideal lateral electric-field distributions between the IDTs. Such spurious responses can be suppressed using various approaches, including placing buslines outside the active region [43], employing buried IDTs [44], optimizing the IDT and film thickness dimensions [45], and introducing quadratic phase variations along the transverse direction through nonuniformly shaped IDTs [46]. In this work, only the placement of buslines outside the active region was adopted.

To synthesize a ladder filter (see the inset of Fig. 3(b)), a frequency shift between series and shunt resonators was introduced, where the parallel resonance ($f_p$) of the shunt resonator and the series resonance ($f_s$) of the series resonator were intended to overlap. The different thicknesses of $LiNbO_3$ used for the shunt and series resonators produce this shift. The mBVD model's $Q$ was assumed to be 50 based on previously measured data and would directly affect the achievable IL and roll-off. Note that $k^2$ was extracted as 47% and 53% for the series and shunt resonators from a fitting using the modified Butterworth Van Dyke (mBVD) model.

The optimized device dimensions achieve a 50 Ω third-order ladder filter. The goal was to produce the lowest possible insertion loss (IL) while maintaining reasonable out-of-band (OoB) rejection of 11.6 dB, within the limitations of the mmWave XBAR platform. As a result, the size contrast between the shunt and series resonators was around 2:1, with both exhibiting relatively compact dimensions. A higher-order ladder configuration and different static capacitances could further improve the OoB rejection. Note that we only considered the thin-film $LiNbO_3$ here for the coupled piezoelectric FEA.

We used a circuit simulator to assess the filter's expected performance using the FEA results. Fig. 3(b) shows a simulated filter passband centered at 21.7 GHz with an IL of 1.72 dB and a 19.9% 3 dB fractional bandwidth (FBW), indicating good performance in this frequency range. Notably, filter performance could be further optimized by accounting for routing and electromagnetic (EM) effects; however, we only focused on this prototype design to benchmark the nonlinearity of mmWave XBAR filters.

## III. Thermal Nonlinear Analysis and Comparison

Earlier research indicated that thermally induced nonlinearity was a critical limiting factor in suspended thin-film piezoelectric resonators, particularly those based on $LiNbO_3$ [39]. In our mmWave XBARs with distributed anchors and piezoelectric film thicknesses on the order of 100 nm, but with lateral dimensions extending to several hundreds of μm (as required for 50 Ω filter matching), the thermal conduction path through the $LiNbO_3$ layer is relatively inefficient. Consequently, within the passband, the filter should conduct high power levels through the resonators, further exacerbating temperature rise and the associated nonlinear behavior. In this work, we performed a comparative thermal FEA of the two platforms to investigate, for the first time, the thermal bottlenecks in mmWave acoustic filters.

COMSOL Multiphysics FEA simulated heat conduction in resonant bodies. The imported device layout was extended in all directions to form a 3D resonator model, including a 5-μm-high air box above the Al IDTs. The resonator body itself was modeled as a generic 100-μW heat source. A thermal insulation boundary condition was applied to all outer faces of the substrate in both the $LiNbO_3$-aSi-$Al_2O_3$ and $LiNbO_3$-aSi-Si platforms. This adiabatic assumption approximates semi-infinite substrate extensions. The remaining boundaries exposed to the surrounding air were maintained at a constant ambient temperature of 20 °C (293.15 K), acting as isothermal heatsinks. To ensure a fair comparison between the platforms, the substrates were extended 40 μm beneath the $LiNbO_3$ film, while the lateral cavity boundaries were positioned 35 μm away from the release windows. Table II lists important material properties used in this study [47]–[49].

[1]Certain commercial equipment, software and/or materials identified here merely specify the experimental procedure. Such identification does not imply recommendation or endorsement, nor that those used are necessarily the best available for the purpose.

TABLE II
MATERIAL PROPERTIES FOR THERMAL SIMULATION

| Materials | Density ($kg/m^3$) | Heat capacity[$] (J/kg·K) | Thermal conductivity[†] (W/m·K) | | |
|---|---|---|---|---|---|
| | | | $k_{11}$ | $k_{22}$ | $k_{33}$ |
| Air[#] | 1.204 | 1,006 | 0.02514 | | |
| Al | 2,700 | 905 | 237 | | |
| $Al_2O_3$ | 3,965 | 730 | 35[‡] | | |
| Z-cut[*] $LiNbO_3$ | 4,700 | 628 | 4.60 | 4.19 | 4.19 |
| Si | 2,320 | 678 | 148 | | |

[#]Air was assumed to follow the ideal gas law, with the reported properties at 20 °C and atmospheric pressure; [*]$LiNbO_3$ was rotated in simulation software to 128° Y-cut; [$]Heat capacity at constant pressure ($C_p$); [†]Off-diagonal components ($k_{ij}$) are zero. [‡]$Al_2O_3$ is weakly anisotropic and is assumed to be isotropic for simplicity in this work.

Notably, heat dissipation in this model can occur through several mechanisms. However, because the device dimensions are relatively large, radiative and convective heat transfer are negligible compared to conductive heat transfer. Heat conduction occurs through three primary pathways: the $LiNbO_3$ film, the aluminum electrodes, and the surrounding air. Because the thermal conductivity of aluminum is more than two orders of magnitude higher than that of $LiNbO_3$, and the metal layer is also significantly thicker, heat conduction through the $LiNbO_3$ film is negligible compared to the air and metal pathways.

The vertical material stack was modeled as a series thermal resistance network, where each thermal resistance could be determined directly from the corresponding material properties. Although geometric discontinuities and variations in lateral dimensions limit the validity of a purely one-dimensional approximation, these effects can be incorporated as spreading thermal resistances [49]. These spreading resistances arise from lateral extensions beyond the finite thermal source and are incorporated into the network. The total thermal resistance of each platform was extracted through steady-state analysis.

Figs. 4(a) and 4(b) compare the effect of air conduction in the two stacks. The major difference between the two stacks is that the $LiNbO_3$-aSi-$Al_2O_3$ device has only a 1 μm air gap beneath the film, whereas the Si substrate is etched by approximately 35 μm due to the isotropic $XeF_2$ etch. The total thermal resistance $R_{th}$ is higher in the $LiNbO_3$-aSi-Si stack; thus, the thermal conductance $G_{th}$ is much lower, with $G_{th}$ = 270 μW/K for $LiNbO_3$-aSi-$Al_2O_3$ and 109 μW/K for $LiNbO_3$-aSi-Si. Following Figs. 4(a) and 4(b), heat dissipates farther through the air gap under the resonator in $LiNbO_3$-aSi-Si than it does through the air gap under the resonator in $LiNbO_3$-aSi-$Al_2O_3$. Due to the low thermal conductivity of air, the larger cavity in the $LiNbO_3$-aSi-Si platform provides greater thermal isolation, leading to a higher resonator temperature rise of 0.92 K compared to 0.37 K in the $LiNbO_3$-aSi-$Al_2O_3$ platform. Additionally, Fig. 4(c) presents a top-down view of heat conduction profile, highlighting the preferential directions of heat dissipation primarily through the Al IDTs into the surrounding ambient air region, with elevated temperatures concentrated within the active region, as expected.

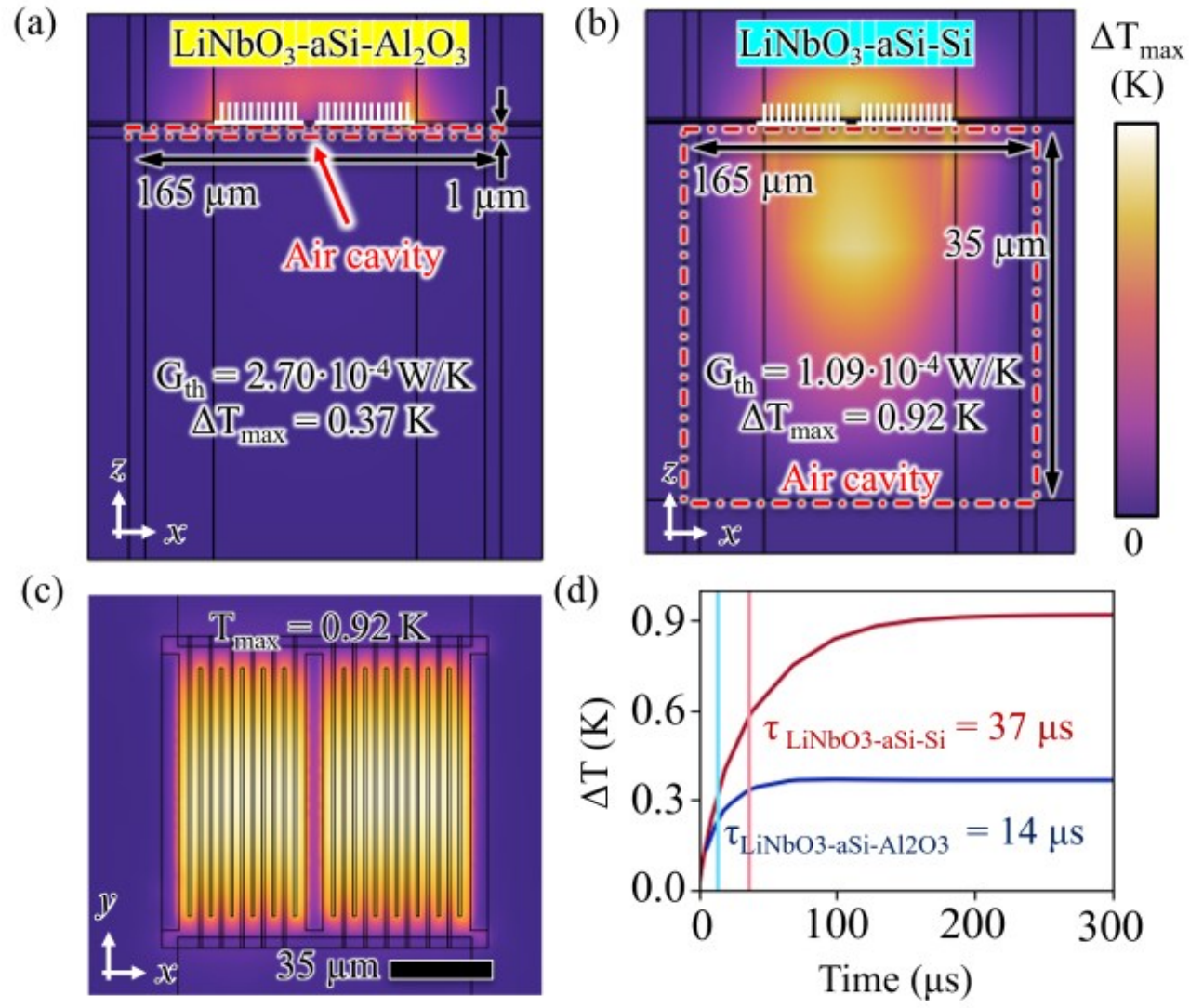

**Fig. 4.** Thermal FEA of released $LiNbO_3$ XBARs comparing heat dissipation on (a) $LiNbO_3$-aSi-$Al_2O_3$ and (b) $LiNbO_3$-aSi-Si platforms under a 100 μW localized heat source in the resonant body, with dashed lines indicating their air cavity region. The thicker air cavity in (b) raises thermal resistance, yielding lower thermal conductance ($G_{th,\ LiNbO3\text{-}aSi\text{-}Si}$ ≈ 109 μW/K vs. $G_{th,\ LiNbO3\text{-}aSi\text{-}Al2O3}$ ≈ 270 μW/K) and a higher peak temperature for the same power. Note that the vertical dimension is not to scale and is exaggerated for visibility. (c) Top-down temperature map highlighting dominant heat-spreading paths through Al IDTs. (d) Thermal step response showing a longer time constant in $LiNbO_3$-aSi-Si ($\tau_{LiNbO3\text{-}aSi\text{-}Si}$ ≈ 37 μs) than in $LiNbO_3$-aSi-$Al_2O_3$ ($\tau_{LiNbO3\text{-}aSi\text{-}Al2O3}$ ≈ 14 μs) platform. Green dashed lines show air gaps of both devices.

Fig. 4(d) shows the thermal step response of the resonators obtained from transient thermal analysis, representing temporal temperature change under a unit step heat dissipation from the device (100 μW in this study). Because both devices were the same size, we expected the similar thermal capacitance $C$, and thus the thermal time constant $\tau = C/G$ to scale inversely with thermal conduction. Note that this thermal time constant indicates how quickly a system responds to a change in temperature. As expected, the results show that $LiNbO_3$-aSi-Si device holds more heat, with much higher final temperature of 0.92 K compared to 0.37 K in the $LiNbO_3$-aSi-$Al_2O_3$ counterpart, but dissipates more slowly, with a proportionally larger time constant of 37 μs compared to 14 μs in the $LiNbO_3$-aSi-$Al_2O_3$. Notably, the steady-state and transient thermal analyses produced consistent results, supporting the validity of the modeling assumptions and simulation setup. Collectively, these results suggest that the $LiNbO_3$-aSi-$Al_2O_3$ is likely to exhibit better thermally limited linearity and less pronounced IMD3.

## IV. MATERIAL ANALYSIS

We characterized 128° Y-cut $LiNbO_3$-aSi-$Al_2O_3$ and 128° Y-cut $LiNbO_3$-aSi-Si stacks provided by NGK Corporation with high-resolution X-ray diffraction (HRXRD) and transmission electron microscopy (TEM). A Bruker-JV D1 X-ray diffractometer in triple-axis diffraction (TAD) mode measured HRXRD and the rocking curves [Fig. 5(a)], which quantify lattice tilt and mosaicity. The $LiNbO_3$-aSi-$Al_2O_3$ sample exhibits a narrow peak with a full width at half maximum (FWHM) of approximately 50” (arcseconds), while the $LiNbO_3$-aSi-Si sample shows a broader peak with approximately 210” FWHM. Both measurements are comparable to commercially available thin-film 128° Y-cut $LiNbO_3$ on $Al_2O_3$ and Si substrates, respectively. Notably, a broader rocking curve indicates increased tilt misorientation and mosaicity in the film.

A FEI Nova 600 DualBeam focused-ion-beam system was used for TEM analysis. Bright-field scanning transmission electron microscopy (BF-STEM) imaging was subsequently performed using a FEI Titan S/TEM operating at 300 keV. Cross-sectional BF-STEM images, shown in Fig. 5(b) and (c) for the $LiNbO_3$-aSi-$Al_2O_3$ and $LiNbO_3$-aSi-Si stacks, respectively, confirm uniform, fully bonded layers in both samples, with no gaps or voids detected at the interfaces.

In addition to crystallinity, thermal compatibility is a crucial factor influencing crystal quality. The calculated in-plane coefficient of thermal expansion (CTE) values at 75 °C for 128° Y-cut $LiNbO_3$, sapphire, and Si are shown in Fig. 5(d). The CTE values increase from $14.4 \times 10^{-6}$ $K^{-1}$ to $15.11 \times 10^{-6}$ $K^{-1}$ [50] for the 128° Y-cut $LiNbO_3$ along the [2110], from $2.55\times10^{-6}$ $K^{-1}$ to $2.92\times10^{-6}$ $K^{-1}$ [51] along all directions for Si, and from $7.09 \times 10^{-6}$ $K^{-1}$ to $7.11 \times 10^{-6}$ $K^{-1}$ [52] along all in-plane directions for c-plane $Al_2O_3$ with a 50 °C temperature rise from room temperature. There is in-plane thermal strain caused by CTE differences between the $LiNbO_3$ film and the substrates (Fig. 5(e)). For the $Al_2O_3$ substrate, we calculated that the maximum thermal strain was 430 ppm for the $LiNbO_3$ film on the $Al_2O_3$ substrate. Meanwhile, we calculated a value of 675 ppm for the $LiNbO_3$ film on the Si substrate. The $LiNbO_3$-aSi-$Al_2O_3$ stack has better CTE matching, which reduces thermal stress during high-power operation. In contrast, the CTE mismatch between $LiNbO_3$ and Si is significantly greater, potentially introducing mechanical strain or warping during fabrication and testing. Combined with the larger release cavity in the $LiNbO_3$-aSi-Si stack, these factors result in reduced thermal dissipation and increased susceptibility to self-heating at elevated RF power.

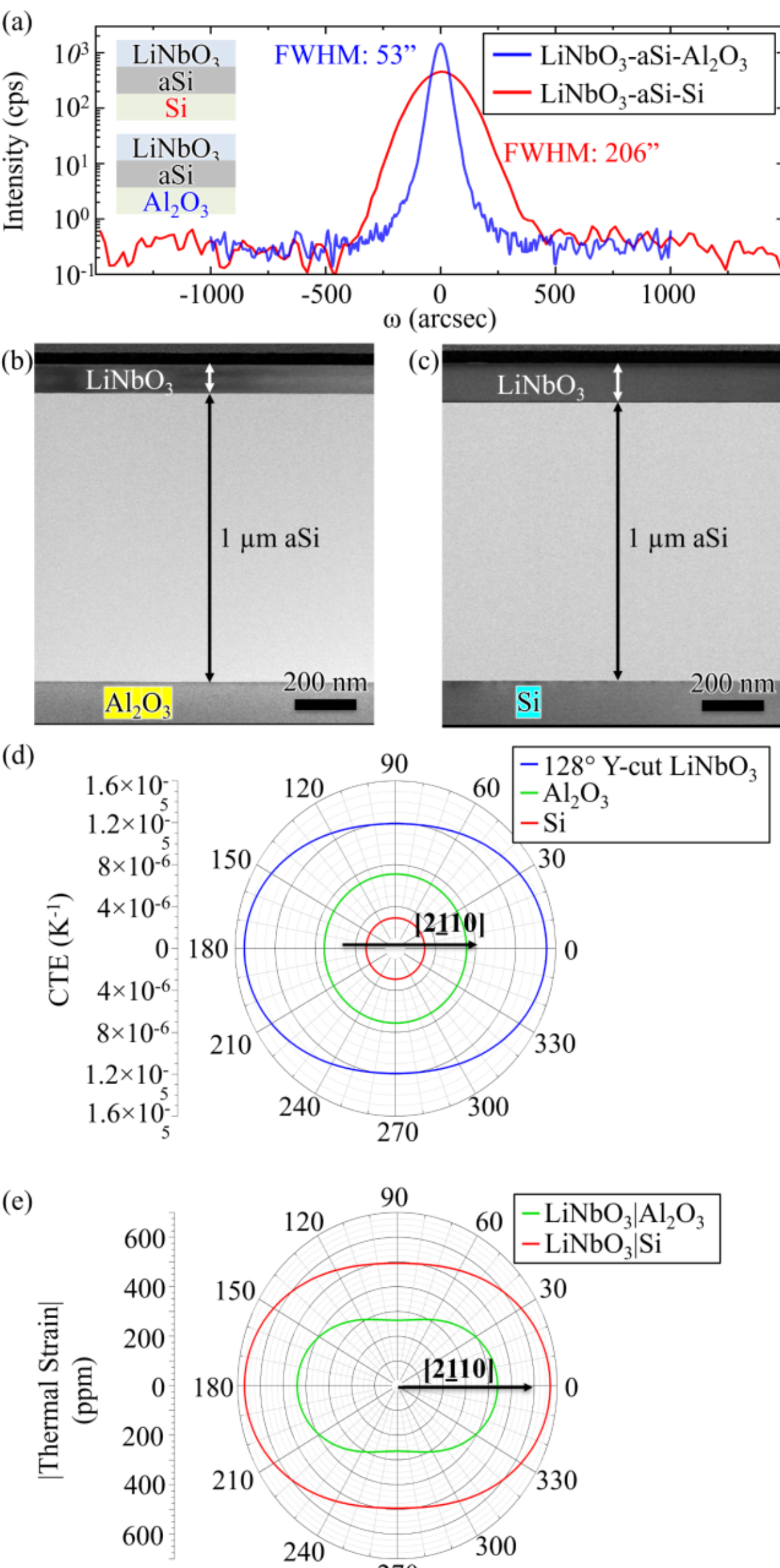

**Fig. 5.** (a) TAD symmetric (0114) rocking curves of the 128° Y-cut $LiNbO_3$-aSi-$Al_2O_3$ (blue) and 128° Y-cut $LiNbO_3$-aSi-Si (red). Cross-sectional BF-STEM images of (b) 128° Y-cut $LiNbO_3$-aSi-$Al_2O_3$ and (c) 128° $LiNbO_3$-aSi-Si. (d) Calculated in-plane CTE at 75 °C for the 128° Y-cut $LiNbO_3$ (blue), sapphire (green) and Si (red), and (e) In-plane strain comparison between the two structures for a 50 °C temperature increase from room temperature up to 75 °C.

## V. Fabrication and Linear Measurements

The fabrication process began by reducing the $LiNbO_3$ sample thickness to 90 nm, which serves as the base thickness for the shunt resonators. Thickness trimming was performed using an ion milling process [53], as it has demonstrated good crystal preservation. Then, we patterned the etch window with lithography and etched into the aSi layer. The long lateral etch windows divide the devices into resonator banks, expediting the release process. Next, local regions for placement of the series resonators were defined using lithography. Afterward, a second round of ion milling was performed on these exposed local trimming regions to achieve a target thickness of 75 nm. The step-height difference between these regions was accurately monitored using atomic force microscopy (AFM), yielding a measured height of 15±1 nm. Electron beam lithography (EBL) was used to pattern the fine features of the metal layer, and Al was evaporated for the metal deposition. The devices were then released by selectively etching the aSi intermediate layer using $XeF_2$. Figs. 6(a), 6(b), 7(a), 7(b), 8(a) and 8(b) show optical images of the standalone shunt and series resonators in the $LiNbO_3$-aSi-$Al_2O_3$ stack, standalone shunt and series resonators in the $LiNbO_3$-aSi-Si stack, and filters in both stacks, respectively. Notably, the released regions of the resonators in the $LiNbO_3$-aSi-$Al_2O_3$ stack extended further than those in the $LiNbO_3$-aSi-Si platform because of a faster etch rate due to less etchable Si volume in the first stack. Moreover, a more uneven etched profile and rainbow-like patterns in the $LiNbO_3$-aSi-$Al_2O_3$ stack are due to induced stress between its thin material layers.

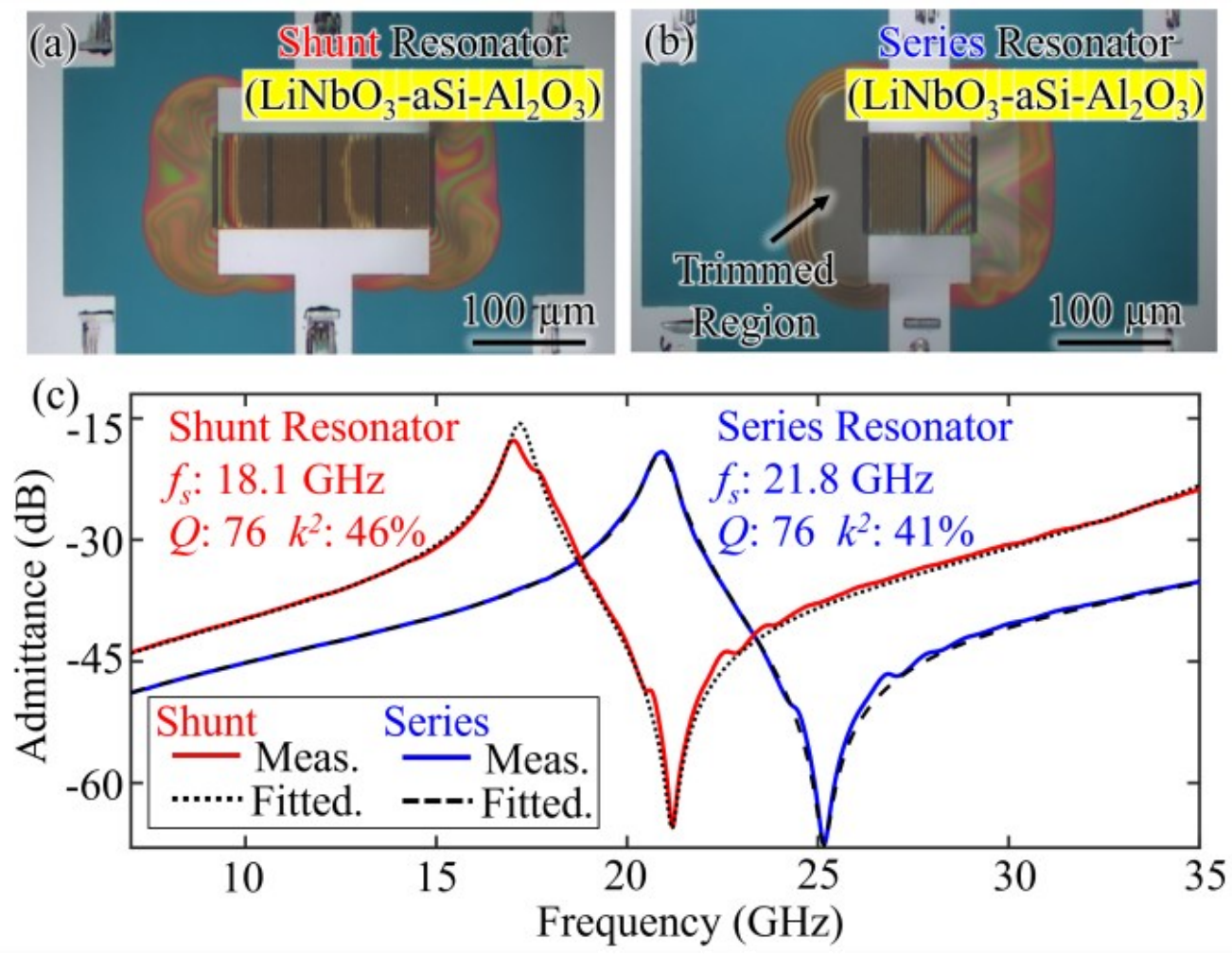

**Fig. 6.** Fabricated (a) shunt (b) series resonators and (c) measured admittance results in the $LiNbO_3$-aSi-$Al_2O_3$ stack.

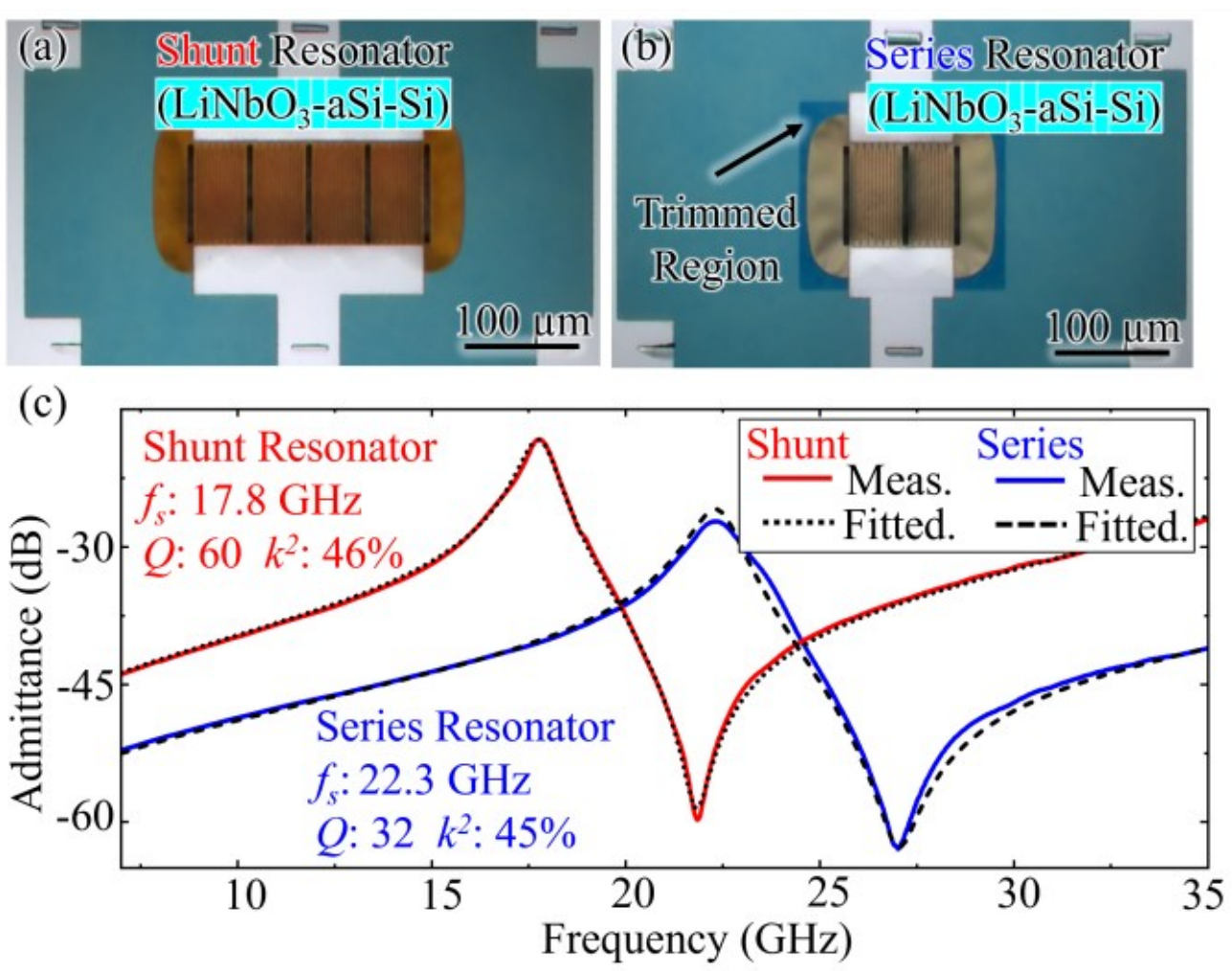

**Fig. 7.** Fabricated (a) shunt (b) series resonators and (c) measured admittance results in the $LiNbO_3$-aSi-Si stack.

The standalone resonators were measured at room temperature using a 67 GHz Keysight vector network analyzer (VNA) with a default VNA power level of −10 dBm. In addition, the measurements are performed after commercial on-chip SOLT calibration and with coaxial cables connecting the VNA and probes. The measured admittance shows the typical response for the resonators fabricated on the $LiNbO_3$-aSi-$Al_2O_3$ stack [Fig. 6(c)] and on the $LiNbO_3$-aSi-Si platform [Fig. 7(c)]. The local trimming to adjust the film thicknesses shifts the mode frequency, leading to differences in resonance frequencies. We used the mBVD model fitting to extract $f_s$, $Q$ and $k^2$ while accounting for parasitic inductance and resistance in the routing. For the $LiNbO_3$-aSi-$Al_2O_3$ platform, the shunt resonator exhibits a resonant frequency of 18.1 GHz, a $Q$ of 76, and an extracted $k^2$ of 46%. The series resonator shows a resonant frequency of 21.8 GHz with a $Q$ of 76 and $k^2$ of 41%. For the $LiNbO_3$-aSi-Si platform, the shunt resonator shows a resonance at 17.8 GHz with $Q$ of 60 and $k^2$ of 46%, while the series resonator has an $f_s$ of 22.3 GHz with a $k^2$ of 45% but a slightly lower $Q$ of 32. The close alignment between $f_p$ of the shunt resonator and $f_s$ of the series resonator near 22 GHz confirms the precision of the local thickness trimming. Following the analysis in Section IV, the $LiNbO_3$-aSi-$Al_2O_3$ stack has better crystal quality than the $LiNbO_3$-aSi-Si stack, which we suspect may have attributed to the higher extracted $Q$ values. Nevertheless, this higher $Q$ also results in more spurious modes at the A1 resonance, as seen in Fig. 6(b) and 6(c). The techniques for improving the suppression were discussed in Section II.

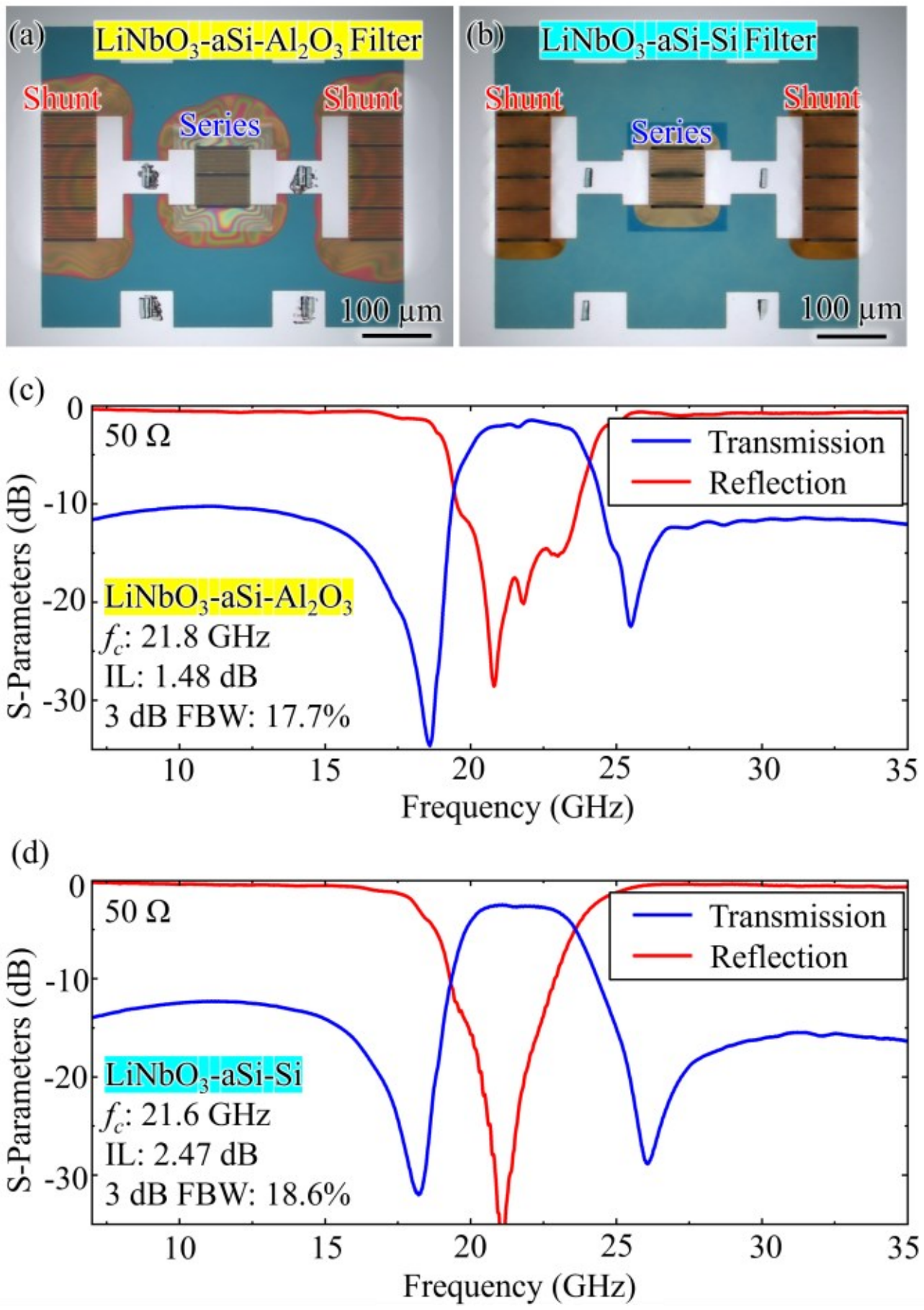

**Fig. 8.** Fabricated filters in (a) $LiNbO_3$-aSi-$Al_2O_3$ and (b) $LiNbO_3$-aSi-Si, with their transmission and reflection responses in (c) and (d), respectively. Note that the data here are all under 50 Ω termination and without post impedance matching. Ports 1 and 2 of the filters are denoted as P1 and P2, respectively.

The filter response for the $LiNbO_3$-aSi-$Al_2O_3$ filter in Fig. 8(c) shows an $f_c$ of 21.8 GHz, an IL of 1.48 dB, and an FBW of 17.7%. In contrast, the filter in the $LiNbO_3$-aSi-Si platform, shown in Fig. 8(d), exhibits an $f_c$ of 21.6 GHz, a slightly larger IL of 2.47 dB, and a 3 dB FBW of 18.6%. These filters have a 50 Ω termination regardless of external impedance matching. The slight frequency shift between the two implementations is primarily due to small thickness variations in the transferred $LiNbO_3$ layer between samples. The measurements are consistent with simulation results [Fig. 3], confirming the efficacy of the local trimming approach for frequency tuning. Moreover, both filters exhibit bandpass with low return loss and well-defined OoB rejection.

## VI. Nonlinear Measurement System Overview

A custom measurement setup characterized the power-dependent behavior of a device under test (DUT) by capturing both its S-parameters and IMD3 in sequence.

The setup, depicted in Fig. 9, configured a four-port 67 GHz VNA for power sweeps of single-tone and two-tone input signals. The setup used all four independent internal microwave sources in the VNA. We used two low-noise amplifiers (>25 dBm output at 1 dB compression) to amplify Ports 1 and 3 and then combine their outputs with a two-way power combiner. Two 20 dB directional couplers sampled the combined two-tone signal to form the first logical port. The reference ($a_1$) and measurement ($b_1$) receivers measured the coupled signals via a jumper on the VNA's front panel. Port 4 served as the second logical port of the measurement system.

To protect the VNA from the high powers used, we connected a 10 dB attenuator to dual 20 dB directional couplers with the coupled signals fed into the reference ($a_4$) and measurement ($b_4$) receivers. 20 dB fixed attenuators kept the receivers from entering compression at high power for the associated coupled signals. We deliberately used very few adapters and short cables downstream of the dual directional couplers to avoid passive intermodulation, which would be indistinguishable from the filter's IMD. The return loss through the power combiner, dual directional couplers, and cable to the coaxial measurement reference plane was sufficiently high (>10 dB) that we did not observe amplifier instability, despite high reflections from the DUT filters at OoB frequencies.

The goal was to measure simultaneous input and output microwave power at the on-wafer reference planes. To calibrate on-wafer power at this reference plane, we adopted a two-tiered S-parameter and power calibration approach. We calibrated the first tier up to the 1.85 mm coaxial connector reference plane before the wafer probes (labelled "Coax Reference Planes" in Fig. 9) with the unknown thru calibration model [54] and a 1.85 mm coaxial calibration kit with polynomial standard definitions. We also performed a power calibration of the VNA receivers at the 1.85 mm reference plane with a thermal power sensor.

After attaching the probes, we performed the second-tier on-wafer S-parameter calibration using a custom multiple-line thru-reflect-line (mTRL) [55], [56] and series-resistor calibration kit [57]. The on-wafer calibration kit had gold coplanar waveguides (CPWs) on c-axis $Al_2O_3$. The mTRL calibration determined the CPW propagation constant ($\gamma$), and the calibration error boxes extended to the center of the thru. Using the estimated CPW $\gamma$, we translated the second-tier reference planes back to the tips of the probes. The probes landed at the very edges of the CPWs, and the reference planes translated to leave 20 µm of CPW (slightly more than the length where probe tips landed) in the second-tier error box. These minimized capacitance errors are caused by differences in landing pad geometry between the calibration kit and the DUT resonators and filters.

The on-wafer series resistor calibration served as a verification of calibration and a means to set the on-wafer reference impedance to 50 Ω [58]. From the mTRL-corrected S-parameters of the series resistor standard, we extracted the capacitance per unit length of the CPW ($C_0$). For the low-loss sapphire substrate, $C_0$ was approximately constant across the frequency range, and the characteristic impedance ($Z_0$) of the CPW was $Z_0 \approx \gamma/(i\omega C_0)$. Finally, we computed an impedance transformer to convert the mTRL error boxes from $Z_0$ to 50 Ω reference impedance at the probe tip reference plane. Notably, all microwave power and S-parameter measurements here were corrected to a 50 Ω reference impedance at the probe-tip reference plane.

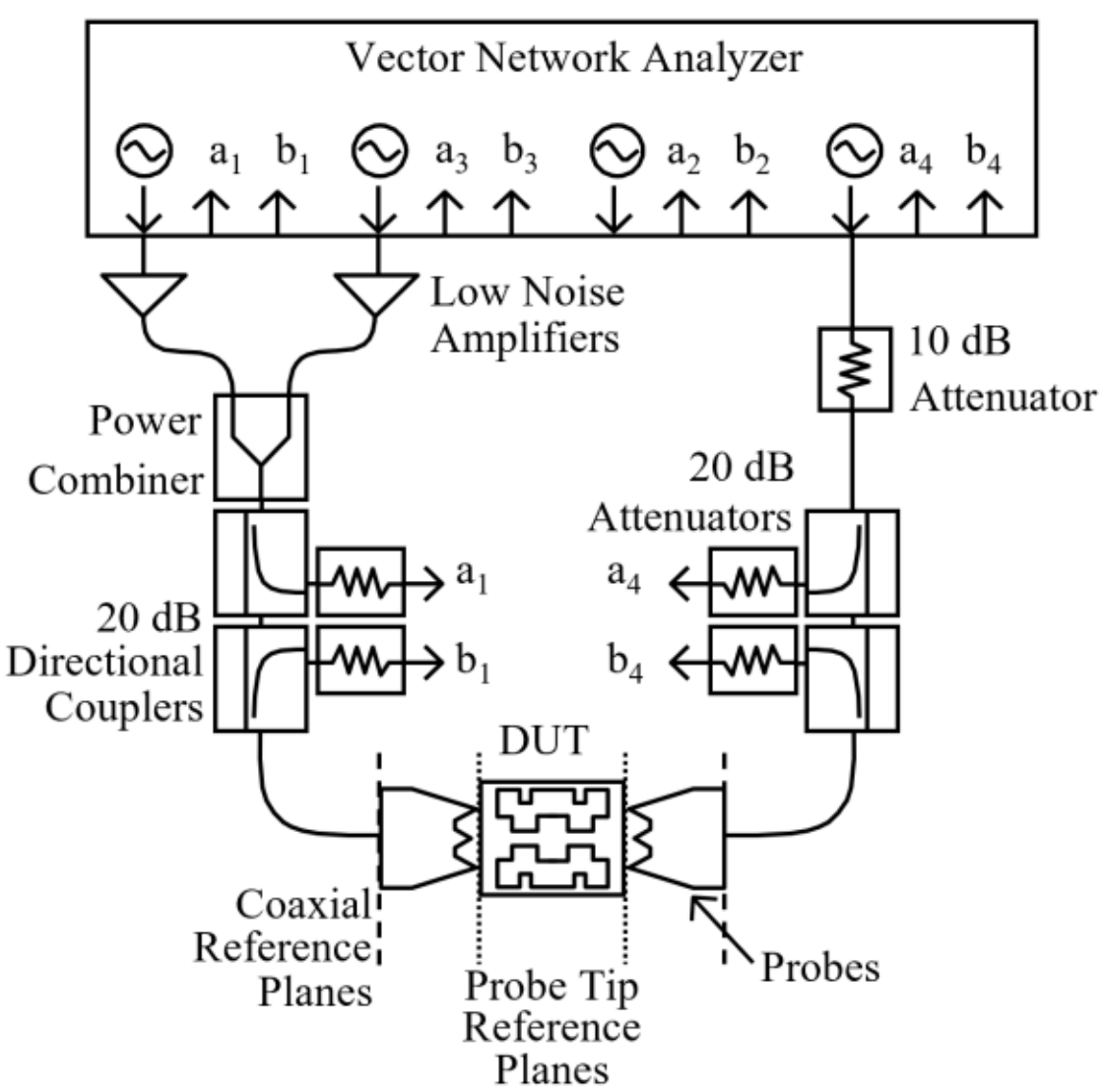

**Fig. 9.** Nonlinear measurement system setup.

We collected all data as raw wave parameters ($a_1$, $b_1$, $a_4$, $b_4$) and processed the first- and second-tier calibrations in the NIST Microwave Uncertainty Framework [59]. For DUT S-parameters, we input single-tone signals and measured at the same frequency. For IMD3, we set the Port 1 and Port 3 sources as generators with a small offset in frequency ($f_1$ and $f_1 + \delta f$) and measured the receivers at the generator and offset IMD3 frequencies ($f_1 + n \times \delta f$, where $n = -1, \ldots, 2$). We chose $\delta f = 1$ MHz to keep the beating of the input RF power faster than the predicted thermal time constants of the filters [60]. $\delta f$ on the order of a few MHz should reveal the intrinsic nonlinearity of the devices for wideband communication signals without spreading the input tones too far apart in the $S_{21}$ response. From the wave parameters translated to the probe tip reference plane, we calculated the switch-term-corrected S-parameters versus input power [61] and the output power at the generator and offset frequencies versus input power for IMD3.

## VII. Power Dependent S-Parameter Measurements

As discussed in Section I, filter distortion is a well-known limitation of acoustic filters when operated under high input power. A particularly important consequence of this behavior is an increase in IL within the passband, as well as frequency drift due to self-heating mechanisms. The ability of a filter to maintain performance across a wide power range depends on several factors, including the material stack properties, the filter's electrical and structural design, packaging, and the design and thermal management of the end-user application (e.g., a smartphone).

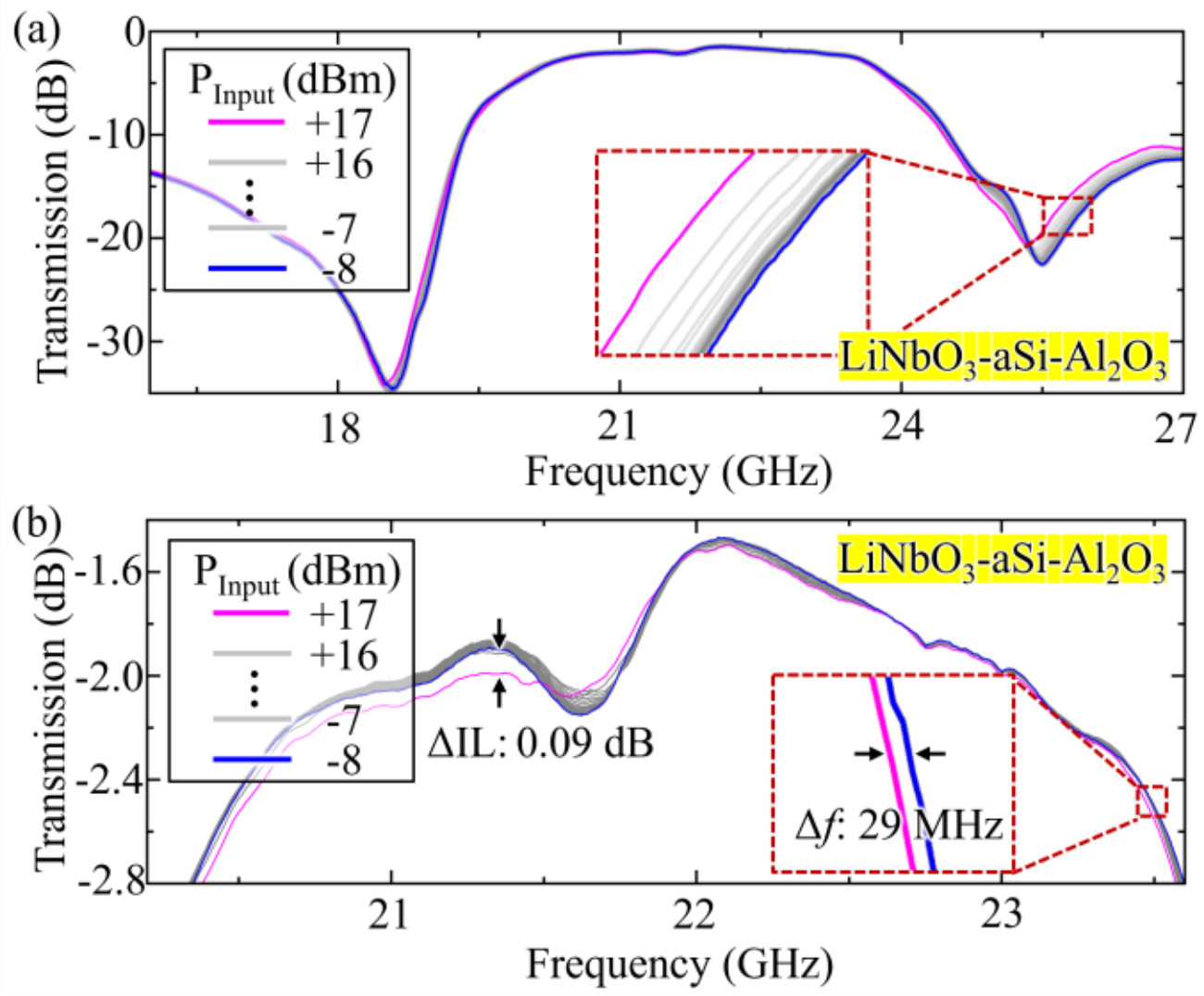

**Fig. 10.** Measured (a) transmission and (b) zoomed-in passband responses of the $LiNbO_3$-aSi-$Al_2O_3$ filter under varying incident power levels. Insets highlight the degradation in ΔIL and $\Delta f$ under high-power excitation.

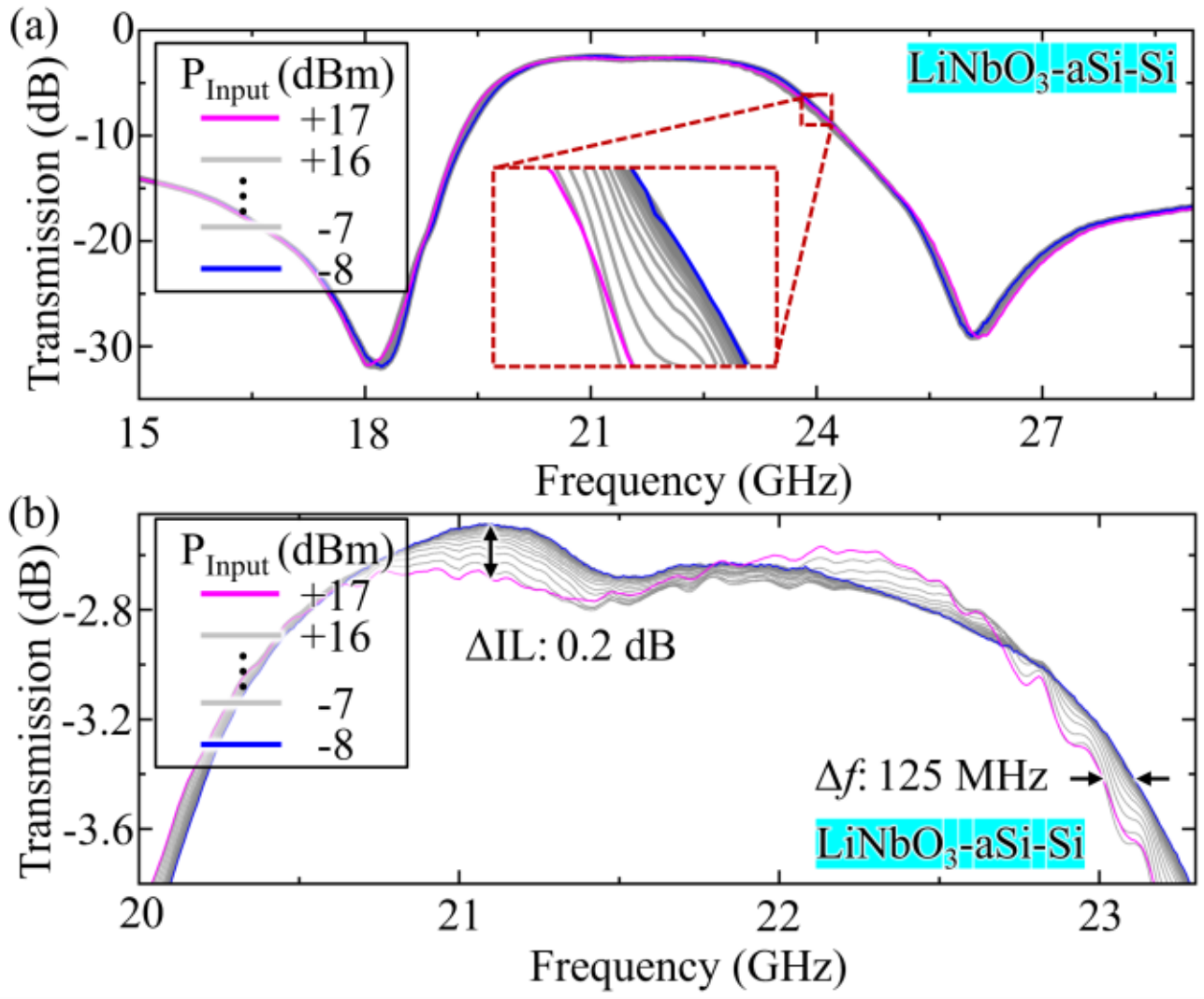

**Fig. 11.** Measured (a) transmission and (b) zoomed-in passband responses of the $LiNbO_3$-aSi-Si filter under varying incident power levels. Insets highlight the degradation in ΔIL and $\Delta f$ under high-power excitation.

For the two platforms presented in this study, we evaluated the IL distortion (ΔIL) at approximately 22.2 GHz, the frequency at which the largest power-level deviation occurred for both devices. We measured the frequency drift ($\Delta f$) at the filter's upper passband edge, 2 dB below the IL minimum, where the change was most significant. Each device was measured across 27 input power levels, ranging from -9 dBm to +17 dBm in 1 dBm increments. The wideband response of the $LiNbO_3$-aSi-$Al_2O_3$ filter is shown in Fig. 10(a), and a zoomed view of its passband in Fig. 10(b) reveals a maximum ΔIL of 0.09 dB and a $\Delta f$ of 29 MHz (1,320 ppm) between the minimum and maximum power levels. By assuming a TCF of approximately −80 ppm/K, the $\Delta f$ of 1,320 ppm corresponds to a temperature change $\Delta T$ of 17 K in the thin-film $LiNbO_3$ layer. In comparison, the $LiNbO_3$-aSi-Si filter, shown in Figs. 11(a) and 11(b), exhibits a maximum ΔIL of 0.2 dB and a $\Delta f$ of up to 125 MHz (5,680 ppm), which corresponds to a $\Delta T$ of 70 K. Such an order-of-magnitude contrast is directionally consistent with $G_{th,\ LiNbO3\text{-}aSi\text{-}Si} \approx 109$ μW/K vs. $G_{th,\ LiNbO3\text{-}aSi\text{-}Al2O3} \approx 270$ μW/K, indicated in Figs. 4(a) and 4(b), indicating accuracy of our modeling approach. These results clearly demonstrate that the $LiNbO_3$-aSi-$Al_2O_3$ platform offers superior thermal stability and linearity compared to the $LiNbO_3$-aSi-Si platform.

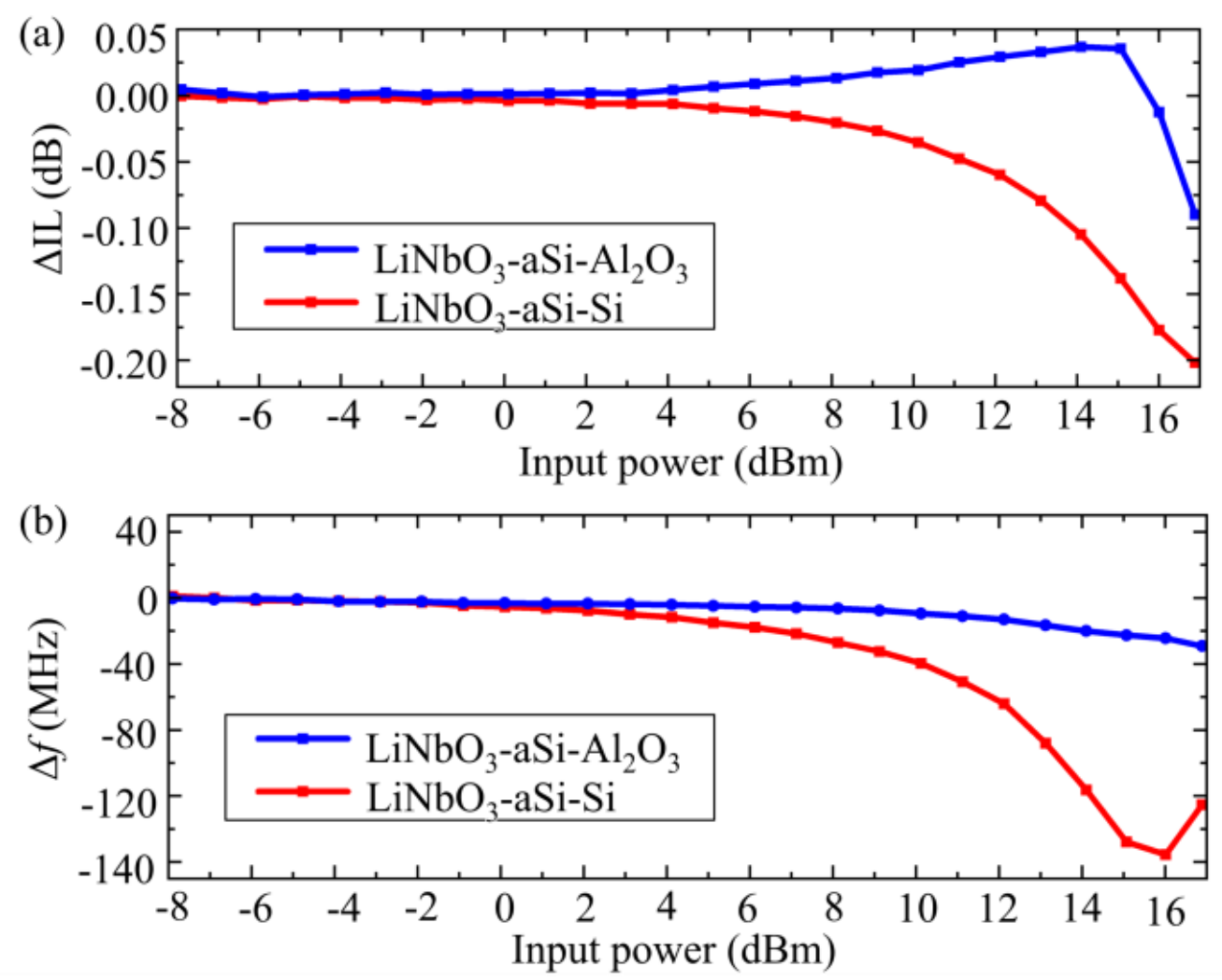

**Fig. 12.** Extracted (a) transmission change (ΔIL), and (b) bandwidth variation ($\Delta f$) versus input power for both platforms.

A comparison of ΔIL, shown in Fig. 12(a), and $\Delta f$, shown in Fig. 12(b), as a function of input power indicates that both devices perform similarly up to +1 dBm, beyond which the $LiNbO_3$-aSi-Si filter degrades more rapidly. The positive ΔIL observed in the $LiNbO_3$-aSi-$Al_2O_3$ filter between +4 dBm and approximately +15 dBm is likely due to a slight initial mismatch between the shunt and series resonators. Initially, frequency drift improves the matching condition, temporarily reducing IL. However, this perceived improvement is short-lived, and the filter performance degrades rapidly thereafter, as expected.

## VIII. Third-Order Intermodulation

Although acoustic resonators are typically passive components modeled using linear equivalent circuits, they have long been recognized to exhibit nonlinear mixing characteristics [62]. In early RFFE applications, these effects were often negligible. However, IMD3 is important in modern communications utilizing higher power and tightly spaced frequency bands as the resulting distortion products can fall within the operating passband, degrading signal integrity.

The IMD3 performance of the filter platforms was evaluated by measuring and computing their IIP3 using the procedure described in Section VI. Two input tones at frequencies $f_1$ and $f_2$, spaced by $\delta f$ = 1 MHz, were swept from −10 dBm to +17 dBm (1 dBm step), and both the fundamental and third-order output signals were recorded. The two-tone measurement was conducted from 17 GHz to

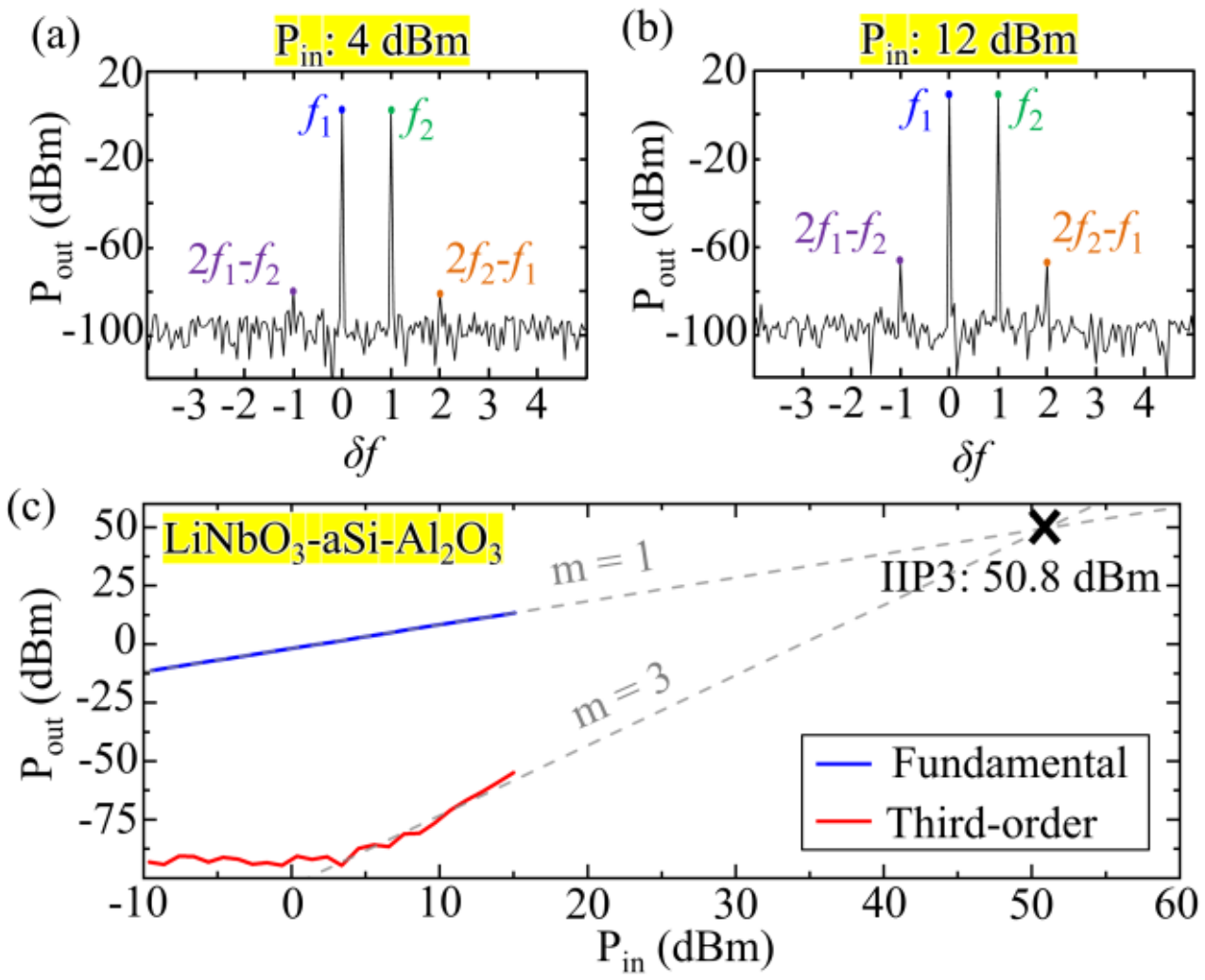

**Fig. 13.** Spectral plots of average output power ($P_{out}$) versus frequency offset from the first input tone $f_1$ = 22 GHz of a filter in $LiNbO_3$-aSi-$Al_2O_3$ at input powers ($P_{in}$) of (a) 4 dBm and (b) 12 dBm. (c) $P_{out}$ versus $P_{in}$ plot showing fundamental and third-order intermodulation components with extracted IIP3.

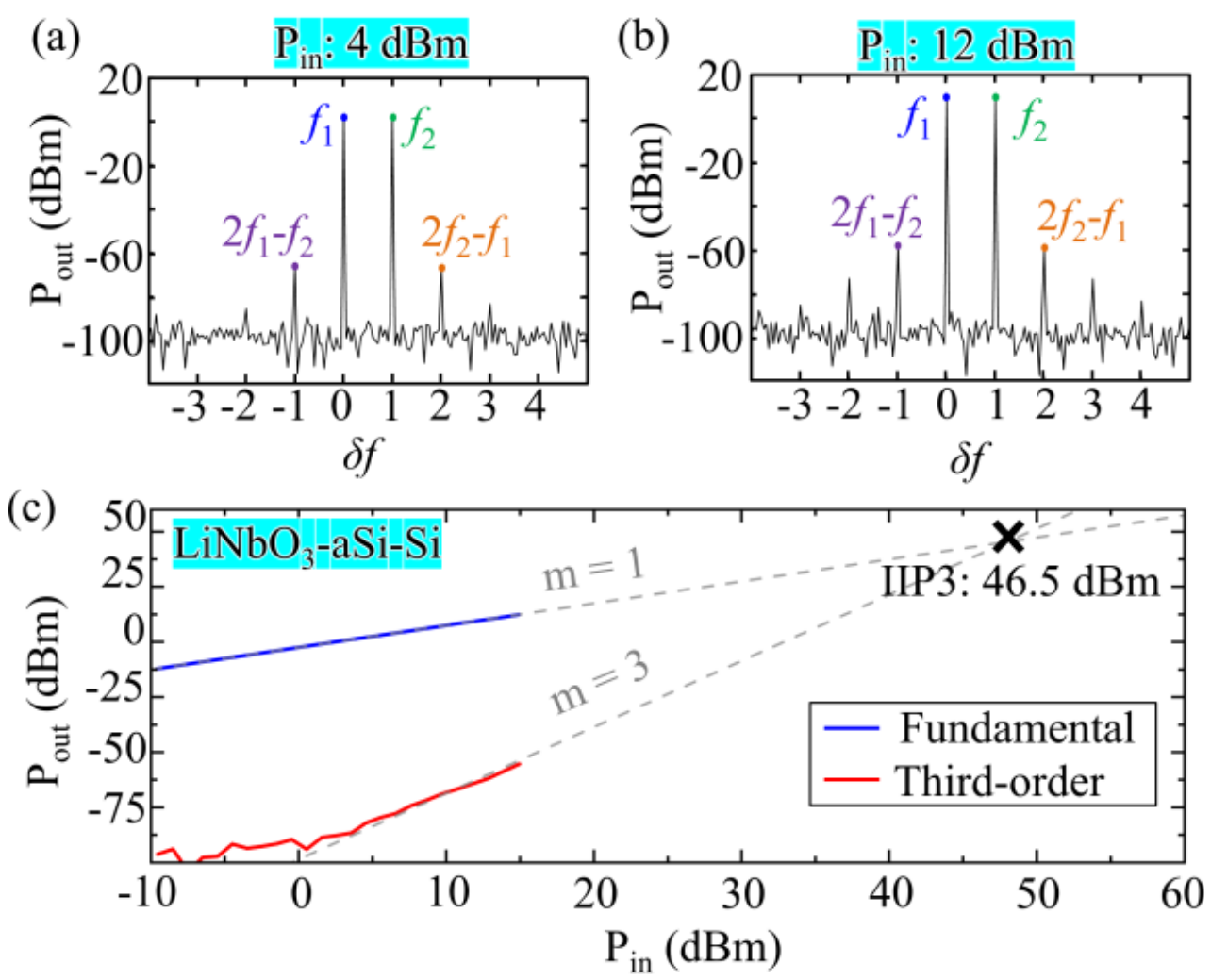

**Fig. 14.** Spectral plots of average output power ($P_{out}$) versus frequency offset from the first input tone $f_1$ = 21 GHz of a filter in $LiNbO_3$-aSi-Si at input powers ($P_{in}$) of (a) 4 dBm and (b) 12 dBm. (c) $P_{out}$ versus $P_{in}$ plot showing fundamental and third-order intermodulation components with extracted IIP3.

28 GHz (0.5 GHz step), covering the entire passband and the adjacent transmission notches on either side. Each frequency point was measured 12 times, and the arithmetic mean was reported to ensure statistical reliability.

At the minimum insertion loss frequency of the $LiNbO_3$-aSi-$Al_2O_3$ filter ($f_1$ = 22 GHz), Figs. 13(a) and 13(b) plot the output power across several $\delta f$ offsets from $f_1$ for the injected tone powers of 4 dBm and 12 dBm, respectively, illustrating the growth of IMD3 products at $2f_1 - f_2$ and $2f_2 - f_1$ with increasing input power. Fig. 13(c) presents the IIP3 extraction at this frequency, yielding a high IIP3 of 50.8 dBm. Similarly, Figs. 14(a) and 14(b) show the output spectra of the $LiNbO_3$-aSi-Si filter with $f_1$ = 21 GHz at its minimum IL frequency, for the injected powers of 4 dBm and 12 dBm, respectively. The IIP3 extraction in Fig. 14(c) indicates that the $LiNbO_3$-aSi-Si filter achieves an IIP3 of 46.5 dBm. These results are consistent with the power-dependent S-parameter measurements in Section VII, confirming the superior linearity of the $LiNbO_3$-aSi-$Al_2O_3$ platform. Notably, stronger nonlinearity in the $LiNbO_3$-aSi-Si platform is also evidenced by the presence of higher-order IMD products, specifically IMD5 at $3f_1 - 2f_2$ and $3f_2 - 2f_1$ visible in Figs. 14(a) and 14(b).

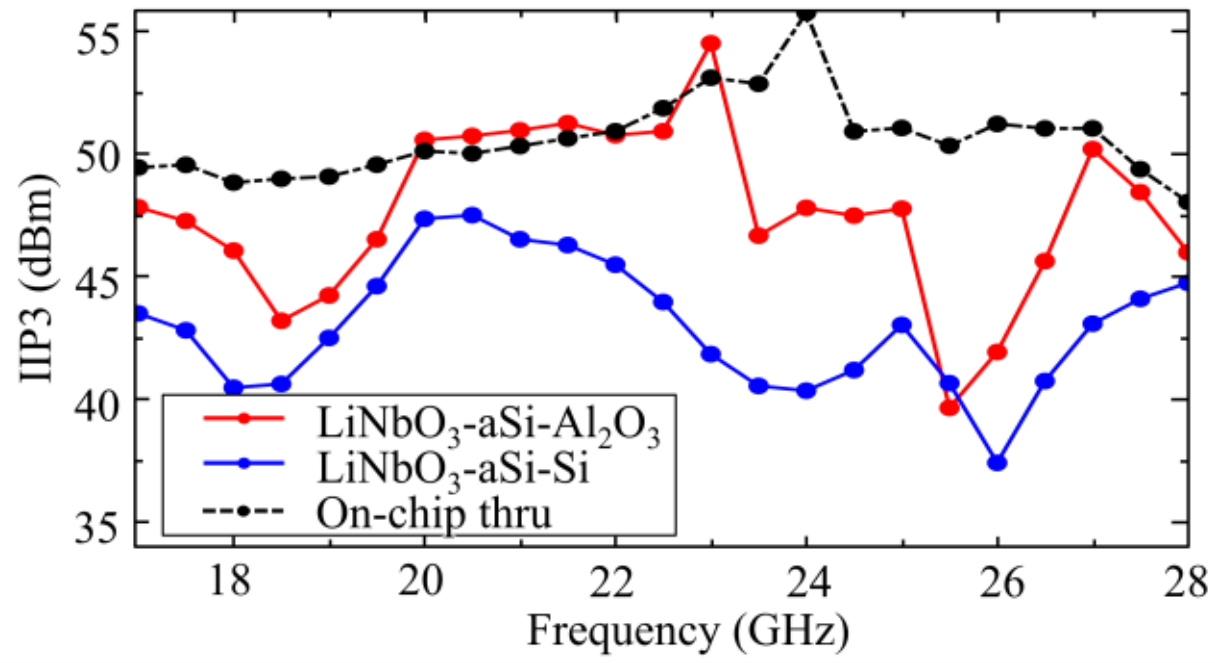

**Fig. 15.** Extracted IIP3 of filters across frequency for both platforms, along with extracted IIP3 of an on-chip thru as a system dynamic range reference, highlighting excellent linearity on both platforms with superior performance in $LiNbO_3$-aSi-$Al_2O_3$.

Considering the input power at minimum IL, the 4.3 dB higher IIP3 of the $LiNbO_3$-aSi-$Al_2O_3$ filter indicates that this device can tolerate approximately 2.69 times more input power, corresponding to 2.7 times higher heat dissipation, than its $LiNbO_3$-aSi-Si counterpart before nonlinearity excessively distorts the output signal. Interestingly, this factor of 2.7 is close to the ratio of 2.5 between the extracted thermal conductances of both platforms (270 μW/K in $LiNbO_3$-aSi-$Al_2O_3$ and 109 μW/K in $LiNbO_3$-aSi-Si), suggesting a strong correlation between our thermally induced nonlinearity assumptions and the measurements. In addition, the slopes of the third-order terms are slightly above and below three for the $LiNbO_3$-aSi-$Al_2O_3$ and $LiNbO_3$-aSi-Si platforms, respectively. Although the origin of this behavior remains unclear, three possible causes are suspected: (1) asymmetric sidebands generated by coupled positive and negative feedback from the constituent shunt and series resonators when driven above or below their respective resonance frequencies [63]; (2) memory effects [64]; (3) limitations in measurement sensitivity. While this phenomenon warrants further investigation, the extrapolated slopes remain approximately three, consistent with the expected third-order behavior.

Fig. 15 plots the IIP3 responses of both filters as a function of frequency, along with the IIP3 of the on-chip thru as a reference for the system's dynamic range. Both platforms exhibit excellent linearity, with IIP3 values exceeding the sensitivity limit of the measurement system across most of the frequency range. This indicates that the actual IIP3 is likely even higher than reported, highlighting the strong potential of these platforms for applications where intermodulation performance is critical. The $LiNbO_3$-aSi-$Al_2O_3$ platform consistently outperforms the $LiNbO_3$-aSi-Si counterpart across nearly the entire frequency range, in agreement with the simulated and measured results presented previously. Furthermore, localized IIP3 reductions are observed near the transmission notches of each filter. The notches below and

TABLE III
SoA Comparisons of Acoustic Filters Beyond 10 GHz

| Ref. | $f_c$ (GHz) | IL (dB) | 3 dB FBW (%) | IIP3 (dBm) | Size ($mm^2$) |
|---|---|---|---|---|---|
| [15] | 17.4 | 1.86 | 3.9 | 36[$] | 0.6 × 0.33 |
| | 17.4 | 3.25 | 3.4 | 40[$] | 0.6 × 0.33 |
| [16] | 11.9 | 1.5 | 6.6 | 36.7[$] | 0.43 × 0.49 |
| [22] | 10.7 | 0.7 | 5.3 | - | - |
| [24] | 19 | 8 | 2.4 | - | 1.2 × 1.2 |
| [32] | 23.8 | 1.52 | 19.4 | - | 0.85 × 0.75 |
| [35] | 50 | 3.32 | 2.9 | - | 0.74 × 0.48 |
| **This work ($LiNbO_3$-aSi-$Al_2O_3$)** | **21.8** | **1.48** | **17.7** | **50.8** | **0.75 × 0.74** |
| **This work ($LiNbO_3$-aSi-Si)** | **21.6** | **2.47** | **18.6** | **46.5** | **0.75 × 0.74** |

[$] The reported IIP3 values were limited by the measurement setup, as discussed in their papers. The actual IIP3 may be higher.

above the passband correspond approximately to the series and parallel resonant frequencies of the shunt and series resonators, respectively, suggesting that the IIP3 at these frequencies is governed by the nonlinearity of the individual constituent resonators.

## IX. Comparison to State of The Art and Discussions

Table III summarizes the current state-of-the-art in acoustic filter technologies operating above 10 GHz. Compared with recent reports in the literature, this work demonstrates that thin-film $LiNbO_3$-based filters combine low IL with excellent frequency scaling, extending to just below the millimeter-wave range. This performance is achieved in a compact footprint of 0.75 mm × 0.74 mm, enabling high-performance resonators without requiring excessively small device dimensions that can complicate the design of 50 Ω-impedance terminals. Furthermore, the 128° Y-cut $LiNbO_3$ provides very high $k^2$, enabling significantly wider 3 dB FBWs of 18%, while preserving excellent linear performance. In particular, the measured IIP3 approaches 50 dBm for both devices evaluated, highlighting the potential of this platform for high-frequency, high-linearity RFFE applications. It should be noted that the performance reported here is indeed limited by the dynamic range of our instrumentation, and the actual IIP3 of the devices may be higher.

Our studies demonstrate that thermal nonlinearity, arising from device geometry and material stack properties, is the primary driver of nonlinear behavior in XBAR filters. Both power-dependent S-parameter and IMD3 measurements show strong correlation with our thermal analysis of XBARs in different $LiNbO_3$ platforms, confirming superior linearity in the $LiNbO_3$-aSi-$Al_2O_3$ platform. The findings also reveal interesting relations between thermal properties and electrical nonlinearity, suggesting avenues for further optimization.

Nevertheless, additional comprehensive studies are needed to establish the causal mechanisms connecting thermal properties and electrical nonlinearity in acoustic devices. Future work will encompass more rigorous thermal modeling through simulation and measurement, IMD3 characterization across a wider dynamic range and varied tone spacings, and investigation of other nonlinear distortions such as second-order harmonics and higher-order intermodulation products.

## X. Conclusion

This work presents a detailed experimental evaluation of XBAR filters operating beyond 20 GHz, characterizing both linear and nonlinear performance through a custom-built measurement setup designed for accurate and repeatable characterization. Thermal modeling and material analysis reveal strong correlations between thermal properties and device nonlinearity, with the $LiNbO_3$-aSi-$Al_2O_3$ platform demonstrating superior linearity. Together, these findings establish thin-film $LiNbO_3$-based filters as a promising technology for high-performance RF applications.

## Acknowledgements

The authors would like to thank Dr. Ben Griffin, Dr. Todd Bauer, and Dr. Zachary Fishman for their helpful discussions.